# On the Challenge of Plasma Heating with the JET Metallic Wall

M-L Mayoral[1,2], V Bobkov[3], A Czarnecka[4], I Day[1], A Ekedahl[5], P Jacquet[1], M Goniche[5], R King[1], K Kirov[1], E Lerche[6], J Mailloux[1], D Van Eester[6], O Asunta[7], C Challis[1], D Ciric[1], J W Coenen[8], L Colas[5], C Giroud[1], M Graham[1], I Jenkins[1], E Joffrin[5], T Jones[1], D King[1], V Kiptily[1], C C Klepper[9], C Maggi[3], R. Maggiora[11], F Marcotte[10], G Matthews[1], D Milanesio[11], I Monakhov[1], M Nightingale[1], R Neu[2,3], J Ongena[6], T Pütterich[3], V Riccardo[1], F Rimini[1], J Strachan[12], E Surrey[1], V Thompson[1], G Van Rooij[13] and JET EFDA contributors[*]

JET-EFDA, Culham Science Centre, Abingdon, OX14 3DB, UK.

[1] Euratom/CCFE Fusion Association, Culham Science Centre, Abingdon, Oxon, OX14 3DB, U.K.
[2] EFDA Close Support Unit, Garching, Germany
[3] Max-Planck-Institut für Plasmaphysik, EURATOM-Assoziation, D-85748 Garching, Germany.
[4] Association Euratom-IPPLM, Hery 23, 01-497 Warsaw, Poland.
[5] Association EURATOM-CEA, CEA/DSM/IRFM, Cadarache 13108 Saint Paul Lez Durance, France.
[6] Association "EURATOM - Belgian State" Laboratory for Plasma Physics Koninklijke Militaire School - Ecole Royale Militaire Renaissancelaan 30 Avenue de la Renaissance B-1000 Brussels Belgium.
[7] VTT Technical Research Centre of Finland, Association EURATOM-Tekes, P.O.Box 1000, FIN-02044 VTT, Finland.
[8] Forschungszentrum Jülich, Institute of Energy Research - Plasma Physics, EURATOM Association, D-52425, Jülich, Germany.
[9] Oak Ridge National Laboratory, Oak Ridge, TN 37831-6169, USA.
[10] École Nationale des Ponts et Chaussées, F77455 Marne-la-Vallée, France.
[11] Associazione EURATOM-ENEA sulla Fusione, Politecnico di Torino, Italy.
[12] PPPL, Princeton University, Princeton, USA.
[13] FOM Institute DIFFER P.O. Box 1207 NL-3430 BE Nieuwegein, The Netherlands.

E-mail address of main author: marie-line.mayoral@efda.org

**Abstract**. The major aspects linked to the use of the JET auxiliary heating systems: NBI, ICRF and LHCD, in the new JET ITER-like wall (JET-ILW) are presented. We show that although there were issues related to the operation of each system, efficient and safe plasma heating was obtained with room for higher power. For the NBI up to 25.7MW was safely injected; issues that had to be tackled were mainly the beam shine-through and beam re-ionisation before its entrance into the plasma. For the ICRF system, 5MW were coupled in L-mode and 4MW in H-mode; the main areas of concern were RF-sheaths related heat loads and impurities production. For the LH, 2.5 MW were delivered without problems; arcing and generation of fast electron beams in front of the launcher that can lead to high heat loads were the keys issues. For each system, an overview will be given of: the main modifications implemented for safe use, their compatibility with the new metallic wall, the differences in behavior compared with the previous carbon wall, with emphasis on heat loads and impurity content in the plasma.

---

[*] See the Appendix of Romanelli F et al 2012 *Proceedings of the 24th IAEA Fusion Energy Conference*, San Diego, US

# 1. Introduction

Following a 20 months shutdown, experiments started again on the JET tokamaks in summer 2012 with a new first wall mimicking the ITER choice of plasma facing material during the active phase i.e. Tungsten (W) in the divertor and Beryllium (Be) in the main chamber. In this contribution, we give an overview of the achievements of the three JET heating systems and of the issues related to their operation in the new JET ITER–like wall (JET-ILW) compared to the JET Carbon wall (JET-C) going through their modifications to be compatible with the metallic environment, the operational changes due to potentially damaging heat loads, their interaction with the plasma facing components (PFCs) and contribution to the source impurities in the plasma. Figure 1 will be used as a guide for the new JET layout. This top view sketches the positions all the heating systems but also of the material used for the different areas. W or W-coated Carbon Fiber Composite (CFC) tiles are highlighted in red. There are situated mainly the divertor areas but also in the main chamber as W-coated tiles were used for the restraint rings (these rings are situated on inner wall with the aim to stiffen the vessel and help keeping the sectors together especially in disruptions) and also as discussed in the next section, for the NBI shine-through areas. Be or Be-Inconel tiles, highlighted in green, were used for the main chamber outer poloidal limiter (PL) and inner wall guard limiter (IWGL), LH and ICRF antenna private limiter and Faraday screens. More detailed description of the JET ITER-Like wall geometry and tiles properties can be found in [1][2].

In the JET tokamak, the plasma heating and current drive is provided by:
- **Neutral Beam Injection (NBI)**. The system consists of two neutral beam injector boxes (NIBs) each equipped with 8 Positive Ion Neutral Injectors (PINIs) [3]. Four PINIs in each NIB are grouped into a 'tangential' bank and four in a 'normal' bank as sketched on Figure 1. Four PINIs in each NIB can be steered between two positions relative to the usual plasma centre (upshifted and standard). The latest NBI system upgrade [4], referred as the Enhancement Program 2 (EP2), was launched in spring 2005 and concluded during the 2011-12 JET experimental campaigns, the first with the new ITER-like wall. The three main goals were: (a) to increase the total injected deuterium neutral beam power from 24 MW to at least 34 MW (with 125 kV / 2.1 MW per PINIs); (b) to increase the NBI pulse duration at maximum power from present 10 s to 20 s and at half power from 20 s to 40 s; (c) to improve the availability and reliability of the NBI system.
- **Ion cyclotron resonance frequency (ICRF) antennas**. The four A2s antennas [5] each consist of four straps and are toroidally spaced around the JET tokamak, as represented on Figure 1. Waves with symmetric spectra ("dipole" phasing; parallel wave number $k_\parallel \sim 6.6$ m$^{-1}$) or asymmetric spectra ("$\pm\pi/2$" phasing, $|k_\parallel| \sim 3.3$ m$^{-1}$) are launched by adjusting the phase difference in between the 4 straps of each antenna. The operating frequency range $f_{ICRF}$ is between 25 MHz and 51 MHz. The record coupled powers, 16.7 MW for 0.3 s (JET pulse 38049 – JET C-wall) and 14.6 MW for 1 s (JET pulse 39960 – JET C-wall), were obtained by operating with small (2-3 cm) limiter - plasma separatrix distance, 51 MHz, and dipole antenna phasing. The issue of operation on the more difficult conditions of ELMy H-mode was solved in the past few years with the implementation of ELM tolerant systems and up to 7 MW was successfully coupled in 2009 on type I ELMy H-mode [6][7]. In C-wall operation (limiter - plasma separatrix distance 5 to 6 cm, fully conditioned antennas), the routine power level was in the 5 MW ($f_{ICRF} \sim 33$ MHz) to 10 MW ($f_{ICRF}>42$ MHz) range in L-mode and 3 MW ($f_{ICRF} \sim 33$ MHz) to 6 MW ($f_{ICRF}>42$ MHz) in H-mode. Note that the ITER-like ICRF antenna [8] was not operated since mid-2009 due to a broken capacitor.
- **Lower hybrid current drive (LHCD).** The launcher operating at 3.7 MHz integrates 48 multi-junctions modules fed by 24 klystrons [9][10]. The usual radiated $n_\parallel$ spectrum is peaked at 1.84 but values between 1.4 and 2.3 can be used. On L-mode plasmas, 7.3 MW (JET pulse 33618 – JET C-wall) has been coupled for 0.2 s and 6.2 MW (JET pulse 34419 – JET C-wall) for 2 s under more steady conditions [11]. In ELMy H-mode, local gas injection can be used to efficiently minimize the reflection coefficient [12] and up to 3.2 MW of LH power could be coupled (JET C-wall) with antenna –plasma distance as large and 15 cm. In C-wall operation (limiter - plasma separatrix distance 5 to 6 cm, fully conditioned launcher), the routine power level was in the 5 MW range in L-mode and 3 MW range in H-mode.

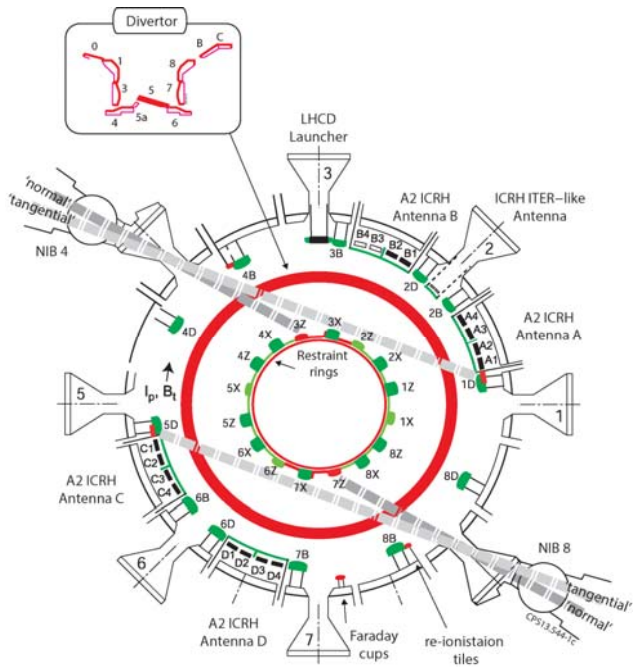

**Figure 1.** Top view of the JET tokamak showing the heating systems layout. Areas with W or W-coated CFC tiles are highlighted in red and area with Be or Be-Inconel tiles are highlighted in green. For more info see [1, 2].

## 2. Neutral Beam Injection

*2.1. Preparation of the JET ITER-like wall in view of Neutral Beam injection*
In parallel with the ITER-like wall installation, the upgrade of the JET NBI system referred as EP2 was completed [4]. All the beam injectors (previously 80kV/52-60A tetrodes PINIs and 130kV/58A triode PINIs) were converted (125kV/65A PINIs). The conversion consisted mainly of an ion source modification from supercup to chequerboard type producing more molecular ions, with higher neutralisation efficiency, and a re-optimisation of the accelerators to increase the beam current. Some beam-line components were also modified to cope with an increase by a factor four in the fractional and molecular residual ion power. In addition the inertial duct liners were replaced by actively cooled copper ones. Finally, several existing high voltage power supply units (160 kV/60A) were replaced with new ones (130 kV/130A) and the power supply layout modified.

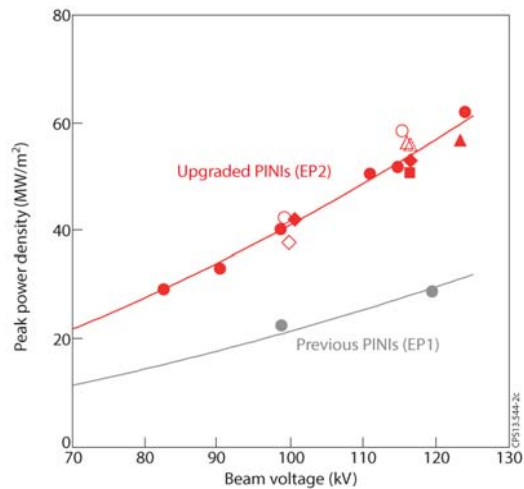

**Figure 2.** Non-attenuated peak power density from D neutral at a 10 m distance (NB Test Bed) for EP1 and EP2 injectors (PINIs).

Because of the increase in the beam center power density and in the pulse length capabilities, particular care was given to the design of the plasma facing components possibly interacting with neutrals injected from the PINIs. Prior to the full use in the ILW environment EP2 injectors were tested on the JET Neutral Beam Test bed and two were installed in 2009 on the tokamak to measure the injected beam power and to confirm that the power loads on various beam-line components were within the predicted margins. These two EP2 injectors were used in support of the JET experimental programme from July to October 2009 and routinely injected around 2MW of deuterium beam power into the JET plasmas, achieving on plasma 112kV, 54A and 9.3s [13]. This power can be compared with the maximum value of 1.5MW obtained with injectors from the previous upgrade and referred as EP1 PINIs. The neutral beam shine-through for the EP2 injectors was inferred from measurements on the test bed and on the inner wall guard limiter (IWGL) made prior to the installation of the ILW of CFC tiles. The experimentally measured shine-trough fractions were found in very good agreement with the ones calculated by the PENCIL code [14].

Non-attenuated peak power densities were obtained on the test bed (10 m distance) as function of the beam voltage both EP1 and EP2 injectors (Figure 2) and then multiplied by 0.8 and 0.4, to take into account the vertical and horizontal angles of the beam trajectories with respect to the JET inner and outer walls respectively. At 125kV, it was then estimated that vacuum peak power densities up to 48 MW/m$^2$ could be expected on the inner wall and up to 24MW/m$^2$ could be expected on the outer wall. From these measurements and shine-through modelling, it was decided to use for areas at risk, W-coated CFC tiles both on the inner wall (with W-coated inner wall guard limiter recessed by 2.5 cm compared to Be ones) and on the outer wall (top and bottom of the ICRF antennas A and C, see Figure 1). An illustration of the possible areas of interaction between the beams and the wall is represented on Figure 3, where the footprint of Oct.8 beams for 'normal' and 'tangential' bank alignment on the inner wall (left) and for 'tangential' bank in the outer wall (right) are drawn.  In addition to the inclusion of W-coated tiles in the main chamber, an upgraded real-time protection referred to as PEWS2 (Plant Enable Window System 2), based on bulk and surface temperature modeling was developed in order to maximize the use of the new NBI power capabilities over a broad range of densities and plasma configurations while staying within the allowed W-coating temperature limit, initially set to 1200°C.

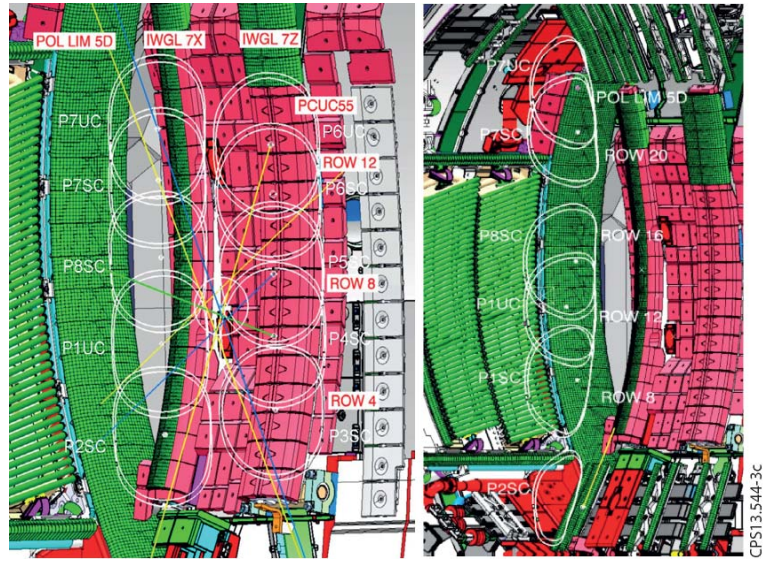

**Figure 3.** Beam footprints on the inner wall for 'normal' and 'tangential' bank (left) and on the outer wall for 'tangential' bank (right). Contours are for power densities of 1 and 0.5 MW/cm². W-coated CFC tiles are in red and Be bulk tiles in green.

Finally, a fraction of the injected neutrals can be ionized in the plasma edge outside the separatrix, drift and then impinge on specific tiles near the outer wall poloidal limiters (PL) 8B and 4B (see Figure 1). These tiles referred to as beam re-ionisation tiles were initially made of CFC. Following estimations (for 8 D beams at 130 kV and used at high density ($1\ 10^{20}$ m$^{-3}$)) [15] that the peak power density on these areas, could be in the range of 5.5 to 23 MW/m$^2$ depending on the limiter - plasma separatrix distance (8 to 5 cm), these tiles were also made of W-coated CFC.

*2.2. Operational experience of the NBI system in the JET ITER-like wall*
A phased approach was taken to increase beam power during the initial operation and for commissioning the new power supplies.  The voltage was slowly increased from at 80 kV ( ~ 1 MW per PINI) to 100kV (~ 1.5 MW per PINI). Once the upgraded real-time protection PEWS2, was commissioned, operation at lower plasma density became possible and as the new power supplies came on line, the injected power could be increased significantly. Consequently in 2012, in JET pulse 83568 an impressive record power of 25.7 MW was obtained at the end of the campaign using 14 PINIs at voltages from 92 to 117kV (Figure 4). Additionally, the averaged coupled power from NIB8 reached its highest value so far (Figure 4) and a record number of pulses with more than 23 MW of NBI power was obtained (Figure 4). Unfortunately technical issues (limitation on stored energy in the power supply transmission lines which required additional inductance) prevented operation at 125kV on plasma before the end of the 2012 campaign.  These issues being now solved higher voltage and power levels are expected for the next JET experimental campaign. In parallel, the new actively cooled duct liners were able to achieve steady-state temperatures (150-200°C) imposing no operational limits. This is shown on Figure 5 for the JET pulse 83307 with 14.6 MW coupled from a single NIB operated with voltages up to 110kV. On this figure, pulse 81511 in which 4 PINIs delivered power for 15 s at voltage of 80 kV, is also shown. This length of pulse was only possible because of the active cooling of the duct liner.

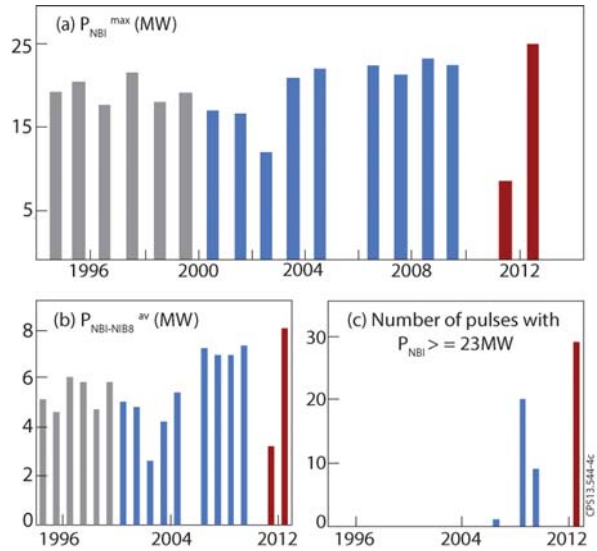 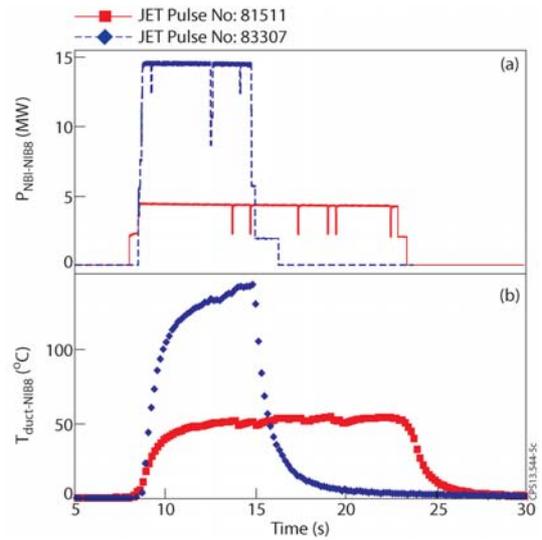

**Figure 4.** Overview of the NBI performance since 1994 with (a) the maximum power coupled, (b) the averaged power coupled from NIB8 and (c) the number of pulses with NBI power above 23 MW. Operation with EP1 upgraded PINIs started in 2000 and with the EP2 upgrade ones in 2011.

**Figure 5.** Time evolution of (a) the NBI power from Oct.8 and (b) related maximum duct temperature.

Hot spots on the beam re-ionisation tiles were observed using one of the cameras from the Protection of the ITER-wall (PIW) [16] viewing system. An example is shown on Figure 6, where localised hot-spots on the re-ionisation tiles adjacent to the outer poloidal limiter PL4B, were seen during a limiter pulse leaning on the outer wall and heated with 5 MW of NBI. The effect of the edge plasma density on the hot-spots on the re-ionisation tiles (for a given radial outer gap (ROG)), is shown on Figure 7 for two ELMy H-modes plasmas. These discharges were performed at the same magnetic field (2.6T), plasma current (1.6MA) and ROG (7cm) and differed only by the gas injection level. In pulse 81560, the density was slowly raised until a hotspot appeared on the re-ionisation tile leading to the termination of the pulse by the protection system. Note that the maximum temperature set by the PIW was at the time very conservative (well below any risk of melting). Future work includes the estimation of the related power densities and improvement of the detection system. Indeed, as most of the main chamber PFCs are made of Be, the emissivity of Be was assumed for the calibration of the cameras used live for protection (PIW) giving an over-estimate for the temperature of these W-coated tiles.

Finally, no increase in impurity levels (expect a minor increase in copper) could be related to the specific application of the NBI power. For example, particular attention was given to the monitoring of any increase in W due to possible sputtering of W-coated CFC tiles by fast neutral deuterium but no evidence of a W source in the main chamber due to NBI was found. This is in agreement with EGDE2D modelling [17] that had shown that the magnitude of the sputtering would not cause significant W radiation. Additionally, prior to the start of the campaign ASCOT simulation [18] were performed in order to check the likelihood to have significant heat loads due to NBI fast ion losses on the outer wall. In the simulations, plasmas with different triangularities were used with NBI power level up to 17.5 MW and beam voltage up to 125 kV. In all cases, only very small losses (<1$^o/_{oo}$) could be predicted with power level lost in the kW range. No problematic fast ion losses were observed during the past campaign.

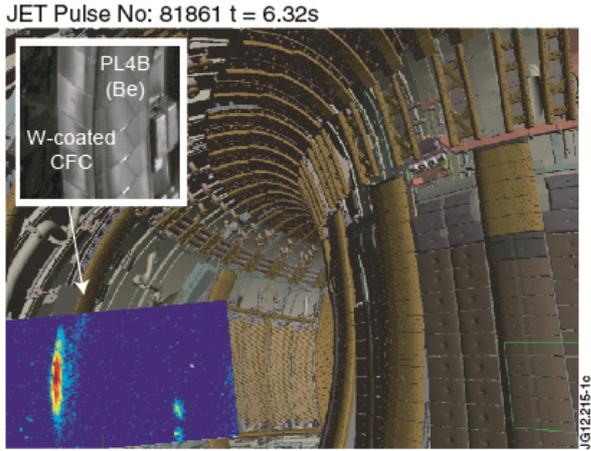
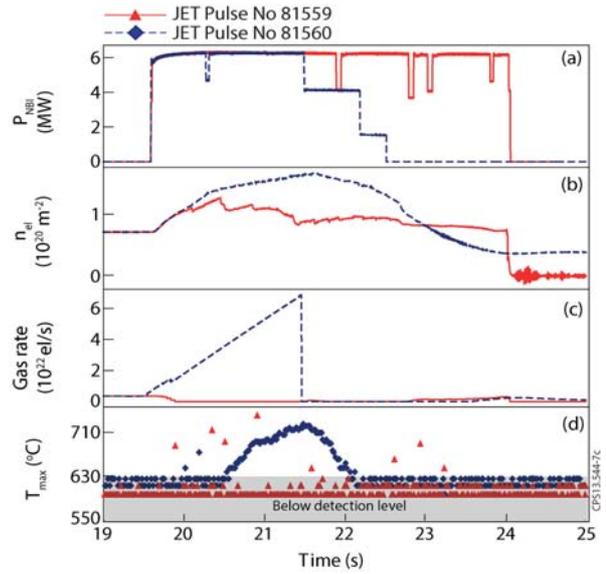

**Figure 6.** Illustration of the heating of the W-coated re-ionisation tiles against PL4B. The IR image originates from an outer limiter pulse with 5 MW of NBI (JET pulse 81861). The maximum temperature hot spot in this pulse using a calibration for W-coated CFC tiles, is 870°.

**Figure 7**. Time evolution of the (a) NBI power, (b) line integrated central density, (c) gas injection levels and (d) maximum temperature using a calibration for W-coated CFC tiles on the re-ionisation tile (near PL4B) for two pulses differing only by the level of gas injected. Note that the camera could not detect the temperature range < 600°C.

# 3. Ion Cyclotron Resonance Frequency Heating

*3.1. Overview of the modifications for ICRF operation with the JET ITER-like wall.* All the plasma facing components around the A2 ICRF antennas were replaced with the exception of the antennas screen bars already made of Be ([19] ; Figure 8 and Figure 9):

- The private limiters on the top and bottom of the A2s antennas consists now of 5 Inconel carriers supporting each 22 bulk Be slices (before one CFC brick per carrier)
- The antenna septum (private vertical limiter in the middle of the A2s antennas), consists now of 5 Inconel carriers supporting each 2 bulk Be bricks (before one CFC brick par carrier). The new septa are recessed by 8 mm to the new main poloidal limiters (4 to 7 mm before) with a slightly modified shape aiming at reducing the thermal loads.
- New flux excluders in Copper (Cu) - coated Inconel have been fitted in between the antennas and the poloidal limiters. Note that the aim of the excluder is to provide a path for the antenna straps mirror current and to avoid high forces at the back of the poloidal limiter during disruptions.

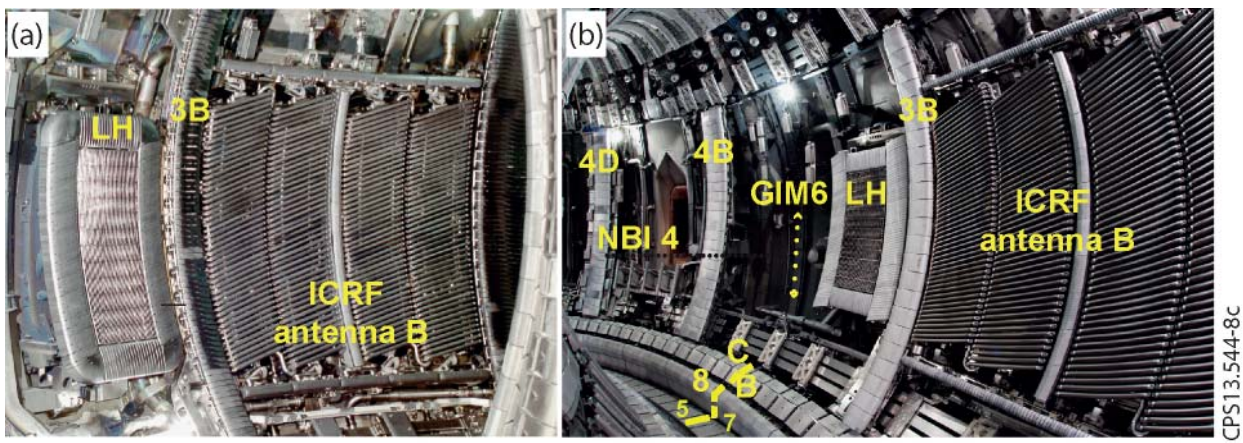

**Figure 8**. Inside view of the JET vessel (a) before 2011; from left to right: LHCD launcher , the narrow limiter 3B, A2 ICRF antenna B and (b) in 2011; from left to right (limiter 4B, the Oct.4 beam duct, limiter 4D, the LHCD launcher, the narrow limiter 3B, the A2 ICRF antenna B). The picture also shows part of the divertor (tile 5, 7, 8, B and C).

During an in-vessel survey after the ILW completion, a misplacement of the flux excluder between the antenna B and the PL3B was observed (see Figure 9). As this excluder was touching the screen bar elements at the bottom of strap 4, it was decided that to avoid possible mechanical damage to the antenna screen mountings during high-current disruptions, no current should flow on the B4 strap until this flux excluder was repositioned. Because of the RF generator configuration this meant not using half of antenna B (i.e. antenna straps B3 and B4) until the next shutdown. Additionally, as B3 and B4 straps were connected to the antenna A straps 3 and 4 (A3 and A4) by 3dBs hybrid splitters in order to couple steady power during the Edge Localised modes (ELMs) [6], some transmission line layout modification were performed in order to keep the full use of antenna A. The transformation was done in time for the experimental campaign starting mid-2011 but left the ICRF system with ½ of antenna B (B3 and B4) not operational, the other half of B (B1 and B2) launching waves with a slightly broader spectra (as operated with 2 straps only) and the ELM–tolerance lost for the A3 and A4 straps. Note that newly designed flux excluder will be fitted during the next shutdown and that the ICRF plant should have its full power capabilities back for the 2013 JET campaign.

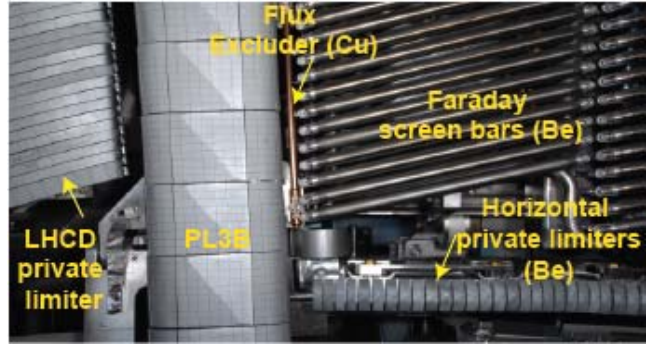

**Figure 9**. Picture of the bottom part of the ICRF antenna B.

*3.2. ICRF heating qualification with JET ITER-like wall*

*3.2.1. Voltage and power achievement.* There was some concern before the first ILW campaign that the modifications of the antennas surrounding structures performed within the ILW project may have had noticeable impact on the antenna performance (reduced electrical strength) and increased the likelihood for arcing. Consequently, during the machine restart, a careful increase of the antenna voltage was performed alternating vacuum conditioning (with or without magnetic field) and plasma conditioning pulses in order to monitor any abnormal behaviour, using μs data acquisition on the ICRF system, visible-light cameras and pressure measurements inside the vacuum transmission lines. Similarly to the system performance in the JET-C, maximum voltages on the transmission lines ($V_{max}$) of ~ 30kV were reached at 33, 42, 47 and 51 MHz without signs of arcing on the new antenna structure.

Using an H minority in D heating scheme, up to 5 MW of ICRF was coupled (JET pulse 81313) using 42 MHz, dipole phasing, a radial outer gap (ROG), of 4 cm. In this pulse, $V_{max}$ was around 25kV, i.e. far from the voltage stand-off limit of the A2 ICRF transmission line (33kV typically). The coupled power being proportional to the square of the voltage for constant coupling, higher power could have being coupled if requested. Nevertheless because of the lowered L to H-mode threshold observed in the JET-ILW compared to JET-C [20], this pulse, as most of the pulses with more than 3 MW, magnetic fields below 2.7 T and electron density below $3.10^{19}$ m$^{-3}$ accessed already H-mode. In type I ELMy H-mode, the maximum coupled power achieved was 4 MW using 42 MHz, dipole phasing, a ROG of 5 cm and $V_{max}$ between 20 and 30kV depending of the transmission line (JET pulse 83398). As explained in the previous section, the capability to couple steady ICRF power, i.e. "the ELM tolerance" was only left for the antennas C, D (still equipped with external conjugate-T [6]) and half of antennas A and B. Taking into account the possibility to adjust the voltages on all the transmission line up to their maximum value (~ 33 kV), 6 MW can be expected when the full 4 antennas will be back. Note that this is slightly lower than the maximum power achieved in the JET-C where 6.5 MW of ICRF power was coupled in very similar H-mode condition (in term of shape, magnetic field, ROG, gas injection, ELM frequency, NBI input power) and for which $V_{max}$ was more in the range of 20 to 25 kV (JET pulse 77404). Close comparison of these two shots showed a lower loading in pulse 83398 (~30%) and hence a higher voltage of the transmission line (that ultimately limits the achievable coupled power). The main reason behind this lower antenna loading is surely the difference in the distance between the antenna straps and the fast wave cut-off density (~ $2.10^{18}$ m$^{-3}$). Indeed, the loading decreasing exponentially as this distance increases [21][22] and the ROG in pulse 83398 was 1 cm larger (5.5 cm) than in pulse 77404. Unfortunately, the uncertainties on the available far scrape-off layer (SOL) density measurements did not allow more precise estimations.

*3.2.2. Heat loads.* A concern when using the ICRF antennas with the new wall, was local heat loads on surrounding Be limiters and antenna private vertical limiters (septum) due to the acceleration of ions in the RF sheath rectified voltages created by the residual parallel electric field $E_{//}$ on the antenna structure [23][24]. This phenomenon was observed previously on Tore Supra [25] but also on JET-C for which estimation of the related maximum heat fluxes were performed, although with the C-wall the work was made difficult because of the presence of layers poorly attached thermally to the bulk tiles and leading to large surface temperature increases [26]. In the recent campaigns, infra-red thermography associated with a thermal modelling of the Be tiles, have enabled further characterisation of these heat fluxes. A detailed description of the method used and related analysis can be found in Jacquet et al. [27]. As represented on Figure 10, the maximum heat-fluxes were found to increase roughly with the electron density at the outer limiter position multiplied by the square of the RF voltage in the transmission lines feeding the antenna (V). Higher heat–fluxes were obtained for asymmetric phasing ($+\pi/2$ or $-\pi/2$) which indeed, are expected to lead to higher $E_{//}$. The hot-spot intensity increased when powering 4 straps (full antenna) compared to two straps (half of an antenna), at constant V. Tentative explanations are changes in the $E_{//}$ field structure, or SOL plasma properties modifications linked to the higher launched power. The highest heat fluxes observed were 4.5 MW/m$^2$ (normal flux on PFCs) when 2 MW/antenna of ICRF power was launched using $-\pi/2$ phasing with a ROG of 4 cm. During the JET-ILW first experimental campaign, the ICRF antenna septa were monitored by the viewing system of the Protection of the ITER-Like wall (PIW) [16]. The design limit before melting for the Be tiles are 6 MW/m$^2$ for 10s; at first, the operational limits were set to ~ 4MW/m$^2$ corresponding to an allowed maximum temperature, measured in real-time by the PIW camera, of 950$^o$C. So far, as typical ICRF pulse durations were below 10s and mainly dipole phasing was used, these RF-sheaths enhanced heat load did not lead to any operational constraint, nevertheless the range of heat-fluxes involved justifies the pursuit of this monitoring.

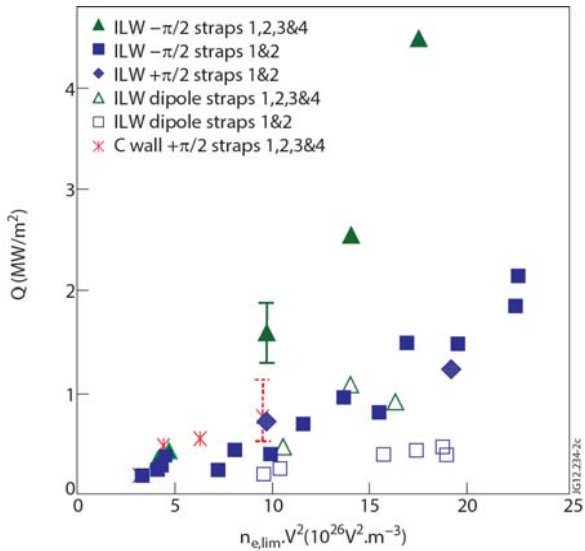

**Figure 10**. Maximum heat-fluxes Q measured around antenna A as a function of $n_{e,lim}V^2$ where $n_{e,lim}$ is the electron density 1 cm in front of the outer PL and V is the ICRF voltage at the antinodes of the transmission lines feeding the straps, averaged over the active straps of antenna A. This figure is reproduced from Jacquet et al. [27]

*3.2.1. Overall plasma properties during ICRF*. So far, only H minority heating in D with the H cyclotron resonance either on-axis or off-axis (using 42 MHz and magnetic field $B_t$ of 2.7 to 2.4T), was used. When the ICRF was used, H was injected to reach levels in the 5% range and ensure good single pass absorption. This heating regime, leads mainly to power deposition on the bulk electrons by collisions with the ICRF - accelerated H ions. A database comparison of the ICRF heated pulses with the JET-ILW (JET pulse range 80661-82240) and with the C-wall (JET pulses range 72150-79853) in similar conditions (no NBI, no LH, 42 MHz, dipole phasing, $B_t$ ~ 2.7 to 2.4T) revealed that the central electron temperature per MW of ICRF power injected (0.5 keV/MW) was slightly lower with the ILW wall but aligned if the higher density used at that time with the JET-ILW, was taken into account. Indeed, at the beginning of the ILW operation the standard operating densities that affect the heating efficiency were generally higher than with the C-wall. An example is shown on Figure 11 and Figure 12, where one can see that additionally to the significant increase in $T_e$ during the ICRF phase, lengthening of the sawtooth periods, characteristic of centrally peaked fast ions pressure [28], was observed. The plasma energy content was found similar with the JET-C and with the JET-ILW (~0.2 MJ/MW) [29]. Additionally as represented on plots (d) of Figure 11 and Figure 12, the plasma energy increase was found slightly higher during 3.5 MW of ICRF heating than during the same level of NBI heating. Note that the signal represented $W_{pl}$ is the total plasma energy derived from magnetic calculation (also called $W_{mhd}$ in the literature) defined as $W_{pl} = W_{th} + 0.75 W_{fast,perp} + 1.5 W_{fast,par}$ with $W_{th}$ the thermal energy, $W_{fast,perp}$ the perpendicular fast ion energy content and $W_{fast,par}$ the parallel fast ion energy content. If the fast ion contribution is removed, similar thermal energies are obtained confirming that the higher radiated power in the ICRF case does not prevent efficient heating.

During ICRF heating, the bulk radiated power was found to increase compared to C-wall operation although not preventing, as just mentioned a significant increase in the plasma energy. Figure 11 and Figure 12 represent the time evolution of two pulses differing only by their density. In these discharges, 3.5 MW of central ICRF heating was applied making the total heating power (with the ohmic power) of 4.8 MW and a matching NBI power phase was added to compare the overall effect of the two heating systems on the plasma properties and particularly on the radiation and impurities levels. Then, one can see that 52 to 56 % of the heating power was radiated during the ICRF phases and 23 to 35% was radiated during the NBI phase, the higher percentages being obtained for the lowest density case. It is interesting to note that generally the radiation from the divertor is as expected because of the lower C content, lower with the ILW than with the C-wall. Nevertheless, during ICRF heating the radiation from the bulk is now higher. The main radiators are W and Ni. The W concentration $c_W$ was estimated [30] both from quasi-continuum (QC) emission at wavelengths around 5 nm corresponding to W ionization stages W27+ to W35+ that have a maximum abundance at $T_e$ ~ 1.5 keV, i.e. in the plasma edge (plot (f) – red plain lines on Figure 11 and Figure 12) and from the l ine radiation of W42+ to W45+ near 6 nm i.e. more centrally at $T_e$ = 2 to 3 keV (plot (f) - black dashed lines on Figure 11 and Figure 12). For pulse 81852 represented on Figure 11, one could see that the W concentration was higher towards the edge ($\rho$~0.5-0.6) in the ICRF case but similar on the plasma centre ($2.10^{-4}$ at $\rho$~0.2) for both the NBI and ICRF heating phases. This slightly hollow W profile for the ICRF and peaked for the NBI pointed to a difference in W source (level and location) and to different edge/core transport effect with the ICRF. It is also worth noting that the moderate increase in plasma density (30%) in pulse 81856 resulted in a reduction on the W concentration by a factor of two (see also section 3.2.2). The increase in Ni observed during the ICRF phase (plot (e) on Figure 11 and

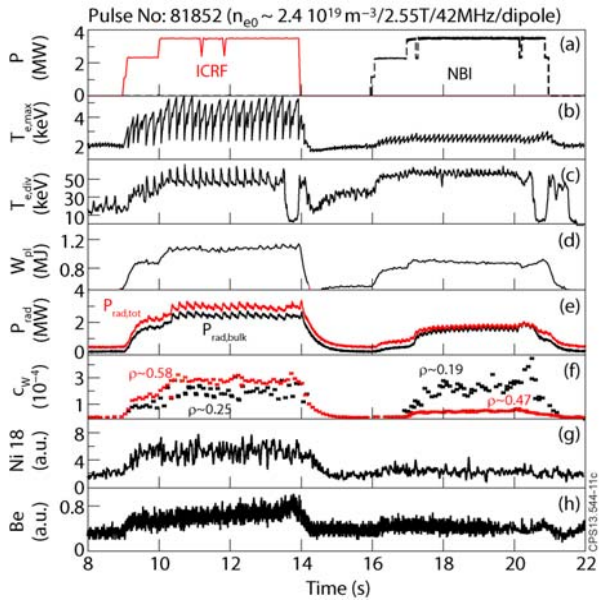 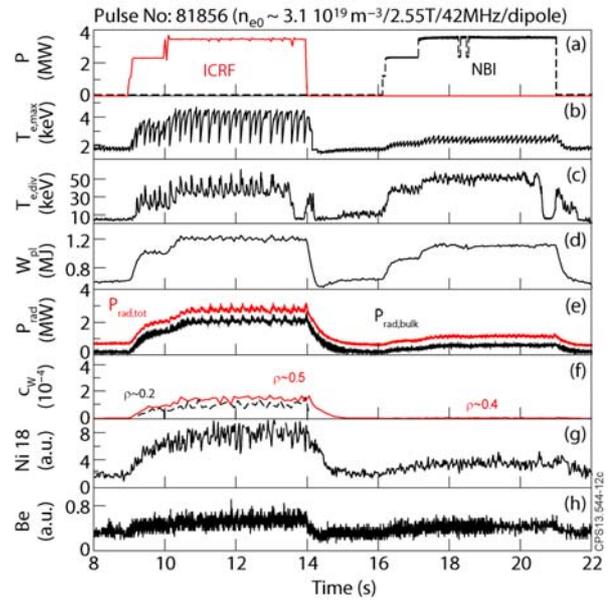

**Figure 11.** Time evolution of (a) ICRF and NBI power (not represented 1.3 MW of ohmic power), (b) central electron temperature,(c) divertor electron temperature (Langmuir probe 17), (d) plasma energy, (e) radiated power (total and plasma bulk), (f) W concentration (see [30]), (g) Ni line intensity (292.00Å line of NiXVIII) and (h) photon flux Be II 527nm (outer divertor). This pulse compared to the one on Figure 12 had a central electron density of $2.4.10^{19}m^{-2}$.

**Figure 12**. Time evolution of (a) ICRF and NBI power (not represented 1.3 MW of ohmic power), (b) central electron temperature,(c) divertor electron temperature (Langmuir probe 17), (d) plasma energy, (e) radiated power (total and plasma bulk), (f) W concentration (see [30]), (g) Ni line intensity (292.00Å line of NiXVIII) and (h) photon flux Be II 527nm (outer divertor). This pulse compared to the one on Figure 11 had a central electron density of $3.1.10^{19}m^{-2}$.

Figure 12) was already present with the C-wall [31] and was found consistent with the values observed previously. With the ILW, it was estimated that the Ni level was contributing the bulk radiation up to a level of 1% [32]. The Be level also increased during ICRF heating. A specific study [33] showed an increase in the Be line intensity during ICRF, for spectroscopic sightlines falling on the faraday screen of the D4 strap and on the poloidal limiter 7B nearby (see Figure 1). Quite interestingly a higher increase in Be from the D4 sightline was observed when the antenna C was in use. A possible explanation for this local increase is an enhanced Be sputtering due to magnetic connection to high RF sheaths potential areas.

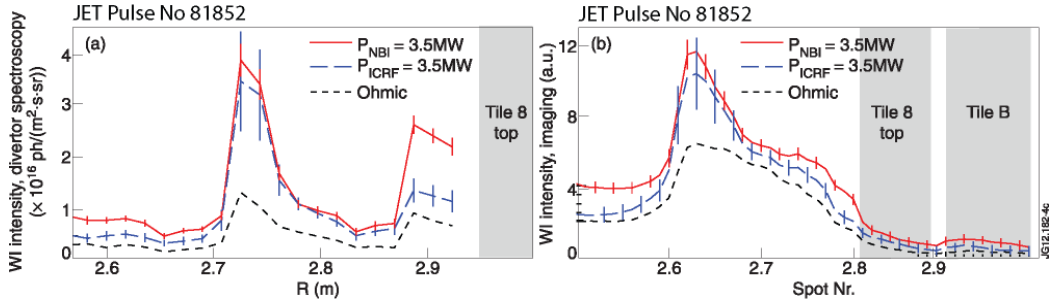

**Figure 13.** WI emission at 400.8 nm from the WI spectroscopy (a) and the WI imaging (b) in #81852 for ohmic only, ICRF and NBI heated phases (3.5MW phases). Top of tile 8 is indicated, which appears larger on the WI imaging system due to viewing geometry. This figure is reproduced from Bobkov et al. [36].

*3.2.2. Heavy impurities sources.* The main questions we tried to answer during the last campaign were: is there an additional W source during ICRF and if yes, where is this source located? The first obvious possible source was the divertor and its entrance (top of tile 8, tiles B and C – see Figure 1 and Figure 8) which can be magnetically connected to the ICRF antennas. Both the visible spectroscopic system [34] measuring the WI (400.9nm) emission from the main divertor area and a camera system [35] with a narrow band filter and an intensifier measuring the WI emission on the outer divertor and its entrance were used to characterised any difference in the W behaviour during ICRF and NBI heating. This detailed analysis [36] was performed for pulse 81852 and the main result is represented on Figure 13: the WI emission representing the W source at the surface, when averaged over the constant heating phase was slightly higher during the NBI phase. Of course as explained in [37], the electron temperature strongly influence the W erosion and the sawtooth activities during the ICRF makes the detailed comparison with the NBI phase challenging. In particular if in averaged the divertor temperature during the NBI is slightly higher than for the ICRF, before sawtooth crashes it becomes lower (plot (c) on Figure 11 and Figure 12). Note also, that Be was identified as the main impurity leading the W sputtering in the divertor [37] and in fact it was shown [36] that at matching $T_e$ points for pulse 81852, the W source was ~ 25% higher during ICRF, compatible with the higher level of Be. It is also important to mention that by varying the choice of: antennas; phasing; and plasma $q_{95}$ the sign of a specific interaction between ICRF antennas and the top of tile 8, was observed (with W fluxes below the $10^{17}m^{-2}s^{-1}$ range) and it is reasonable to assume that divertor entrance tiles C (unfortunately not visible by any diagnostics), might be a source of W.
Interactions with main chamber W-coated tiles (shine-through areas on the inner wall: inner wall guard limiters IWGL 3Z and 7Z, shine-through areas on the outer wall: top of antenna A and D, restraint ring, faraday cups – see Figure 1 and Figure 8) were also considered, particularly as it was found that limiter pulses specifically designed to minimise contact with the divertor, had higher W levels when heated with ICRF compared with NBI. Evidence of an interaction with the main chamber was found by doing overnight Be evaporation as illustrated on Figure 14. Comparing the JET pulse 83383 prior to the Be evaporation with pulse 83428, the first after the evaporation, one could clearly see a strong reduction in the Ni line, W line (the W concentration estimated from SXR analysis [38] showed decrease was estimated to be the 40-50% range) and radiated power (by 45%). Note that the Be layer deposited (~ 3nm) was expected to disappear very quickly (~ 1 pulse) in areas in direct contact with the plasma. The fact that after 11 ELMy H mode pulses, the radiation level in pulse 83442 was still lower than in pulse 83383 prior the Be evaporation, tend to indicate a source of impurities from W-coated and Inconel recessed areas although processes to explain this interaction are still speculative.

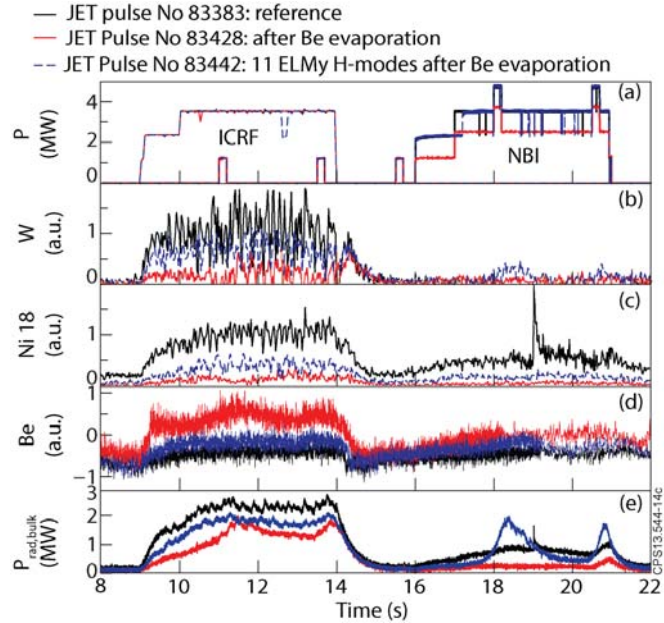

**Figure 14** Time evolution of (a) ICRF and NBI power, (b) W line intensity (mid-plane VUV spectroscopy - integration over the broad feature that could contain also Ni features), (c) Ni line intensity (292.00Å line of NiXVIII from mid-plane VUV spectroscopy), (c) photon flux Be II 527nm (outer divertor) and (e) bulk radiated power. The spike in the Ni line for pulse 83383 is due to Ni laser-blow-off performed at 19s.

*3.2.2. Factors influencing the impurities content.* Experimentally, the first parameter influencing strongly both the W and the Ni level is the edge density. This is shown in details in [36][37] for the W in the JET-ILW, in [32] for the Ni in the JET-ILW but also in [31] for the Ni in the JET-C. A number of different processes can play a role: a) a decrease of the impurity source; b) a change in plasma transport properties degrading impurity confinement; c) a direct dilution of the impurities in the plasma. This effect can be seem when comparing the JET pulses 81856 (Figure 12) and 81852 (Figure 11) that differ only by the density. Pulse 81856, that had an electron density higher (line integrated density at R = 3.9 m was $1.8.10^{19}$ m$^{-2}$) than 81852 (line integrated density at R = 3.9 m was $1.2.10^{19}$ m$^{-2}$) had clearly much lower W, Ni and Be content both in the NBI and the ICRF phase. Another parameter that affects the impurities content is the phasing of the ICRF antenna; with $\pm\pi/2$ leading to higher Ni [32] and higher W emission from the divertor baffle [36]. This effect is illustrated on Figure 15, where the Ni content evaluated for several L-mode pulse of the JET-ILW was found much higher when $-\pi/2$ phasing was used compared to dipole phasing. Finally during a scan of the H levels (H%) in plasma up to 30%, a drop in radiation accompanied by a drop in $c_W$, Ni content and Be flux was observed, with a minimum for H% around 20%. Interestingly, the net heating efficiency only slightly decreased for H% > 20%. As H% was further increased, the radiation rose again, probably due to the strong drop in heating efficiency (see [29]). Although the effect of the H% was remarkable, at the moment no firm conclusion on the processes involved can be drawn because of the concomitant change in the edge density when more H gas was injected.

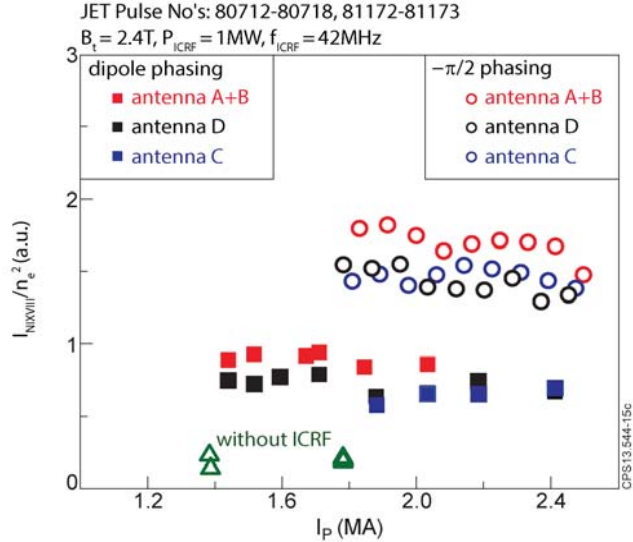

**Figure 15.** Correlation between the Ni content and the plasma current $I_p$ for 2 different antenna phasings. See more in Czarnecka et al. [32].

The production of impurities during ICRF operation with a metallic first wall is not a new phenomenon and was studied in the recent years particularly on AUG [39][40][41] and C-mod [42][43]. The reason leading to this increase is believed to be in most case due to acceleration of light impurities in the RF rectified voltages, enhancing consequently the sputtering of metallic components. Either antennas design improvements and/or coating of the metallic areas connected to these high potential regions, have allowed both in C-mod and AUG, a strong reduction of the impurity content, confirming the role of the RF sheaths. The fact that at JET the phasing is affecting the impurities content goes in the direction of an RF sheaths effect. Also, recent modelling of the A2 antennas with TOPICA code [44] showed that the residual parallel electric field generally present on the antenna structure extends for these antennas even to the surrounding poloidal limiter (see [36]), increasing the possibility of magnetic connection between areas with RF–enhanced voltages and the divertor entrance. It is then more likely that this effect plays a role also in JET although it was not experimentally clearly diagnosed. Nevertheless, to explain the interaction with recessed areas as suggested by the Be evaporation another explanation was needed. The role of "fast" neutrals enhancing the W sputtering was envisaged. Nevertheless, for pulse 81852 a quantification based on neutral particle analyzer measurements for energies of up to 30 keV, showed that neutral H in the ICRF case could lead to very low W flux ($8.7.10^{11}$ $m^{-2}s^{-1}$ range, integrated over the energy range) and that neutral D in the NBI case could lead to W flux of $9.10^{12}$ $m^{-2}s^{-1}$, hence favoring higher W release by charge exchange particles in the case of NBI heating (see [36]).

Finally, in parallel with differences in the W source, differences in transport can be expected for the NBI and ICRF cases. Particularly as ICRF led to strongly peaked electron temperature profiles and strong sawtooth activity. A complete transport analysis of the pulses here presented is outside the scope of this paper nevertheless the following results were obtained so far for pulse 81856 (assuming that the ion temperature $T_i$ is equal to $T_e$, as no $T_i$ measurements were available):
- The European transport simulator (ETS) was used [45] assuming similar impurities transport coefficient for the NBI and ICRF phase (computed from a Bohm-GyroBohm model – analytical description for anomalous transport). The main conclusion being that best agreement with the radiated power and effective charge experimental profiles could be obtained by assuming during the ICRF an increased boundary impurity source (and zero convective velocity) although the effect of a radially shaped convective velocity that was not investigated, could not be excluded.
- Using QuaLikiz [46] (first principle for anomalous transport), an outward diffusion and inward convection both higher for the ICRF than for the NBI was found. For $0.3 < \rho < 0.7$ the inward convection dominated giving a W flux inward and higher for the ICRF. From both the ICRF and the NBI case, the neoclassical diffusion coefficient estimated from [47] was found much smaller than the turbulent diffusion coefficient at all radii and the neoclassical convection smaller than the turbulent one expect for $\rho>0.9$. The conclusion of this set of simulation was that at the edge for both the ICRF and the NBI case both turbulent and neoclassical W convection were inward (with neoclassical dominating) and that more in the core ($0.3<\rho<0.7$) a turbulent inward convection higher for the ICRF dominated. Nevertheless, no results from transport simulations of the plasma core inside $\rho = 0.3$ and taking into account the sawtooth activity is yet available for these pulses and the difference between the peaked profiles for the NBI case and more flatter one for the ICRF, is yet to be modelled.

## 3. Low Hybrid Current Drive

*3.1. Modifications for operation in the ILW.* The main modification to the LHCD launcher (see Figure 8) was the change of the protective frame from CFC slices to Be ones. The grill itself, made of Cu-coated stainless steel, was unchanged. The launcher position was adjusted following photographic measurements and hot spot studies [26] performed before the shutdown and that showed that its position was 19 mm forward of that indicated by the launcher position sensor. The first challenge when applying LHCD with the metallic wall was to avoid the potentially damaging heat flux due to the generation of fast electrons in front of the launcher, that can lead to very localised heat loads on magnetically connected PFCs. This phenomenon was quantified with the C-wall and the maximum heat flux projected onto the tile surface was estimated to ~ $7MW/m^2$ in the JET-C and in worst case conditions [26]. The second challenge was arc detection, which can eventually lead to plasma disruptions due to high impurity influxes [50]. In order to further study the LH related hot-spots, protect Be components and develop new arc detection scheme, a dedicated viewing system (referred to as KL10) consisting of an IR camera, a visible camera and of 4 pyrometers was installed to monitor the LH launcher.

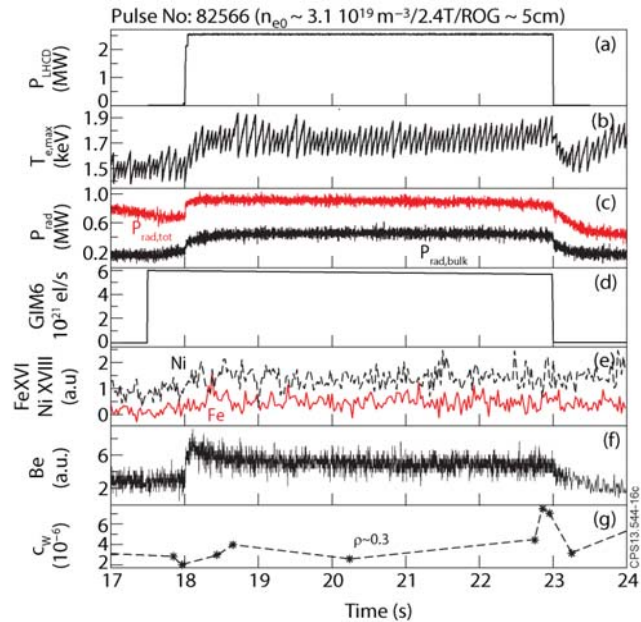

**Figure 16** Time evolution of (a) the LH power, (b) central electron temperature, (c) bulk and total radiated power, (d) $D_2$ gas injected from the gas inlet module 6, Ni and Fe line intensity (from mid-plane VUV spectroscopy), photon flux Be II 527nm (outer divertor) and W concentration from quasi-continuum emission [30].

*3.2. Operational experience with the ILW.* The LH power was gradually increased up to the maximum allowed without full development of dedicated protection viewing systems i.e. 2.5 MW (180kW /per klystrons - $15MW/m^2$ power density). A typical pulse is represented on Figure 16. No noticeable impurity increase could be observed. The edge W concentration was very low ($10^{-6}$ range) and the more central W concentration was below the detection limit. Good coupling conditions as with the C-wall, was ensured by injecting $D_2$ gas from the gas injection module 6, which is a pipe situated near the LH launcher (see Figure 8) and, so far, no degradation compared to the C-wall was observed [48][49]. Unfortunately, due to delays in the KL10 viewing system commissioning, higher power could not be reached during the first ILW campaign. Nevertheless, by the end of the campaign, first IR camera observations were obtained although absolute temperature values will only be available in the next campaigns. This will allow proper documentation of launcher structure heating and fast electron hot-spots characterisation on poloidal outer limiter PL3B. Note that in the conditions allowed so far, none of the LH related hot spots observed by the PIW were of a concern for the wall integrity. The first images from the visible camera highlighted the necessity to filter out visible light above 500nm to reduce the $D_\alpha$ line emission during $D_2$ injection from the gas injection module 6. A filter was also installed to remove Be lines. On Figure 17, we can see images from this new visible camera during an arc propagating on the LH launcher. Although the arc is detected by the existing protection system (based on reflected power imbalance and impurity radiation from the bolometers) [50] and the LH power switch-off at 21.3s, one can see that the arc is visible by the camera 20ms before. It is likely that if this arc could have been extinguished earlier, its propagation and associated Fe influx would have been avoided. So far 16 arc cases were seen by the camera (in 231 pulses). Most of the arcs were stopped by the existing arc detection system but 4 were not stopped in time and led to plasma disruptions due to large Fe influxes. Development of a real-time arc detection system using bright spot detection which will complement the existing system is on-going.

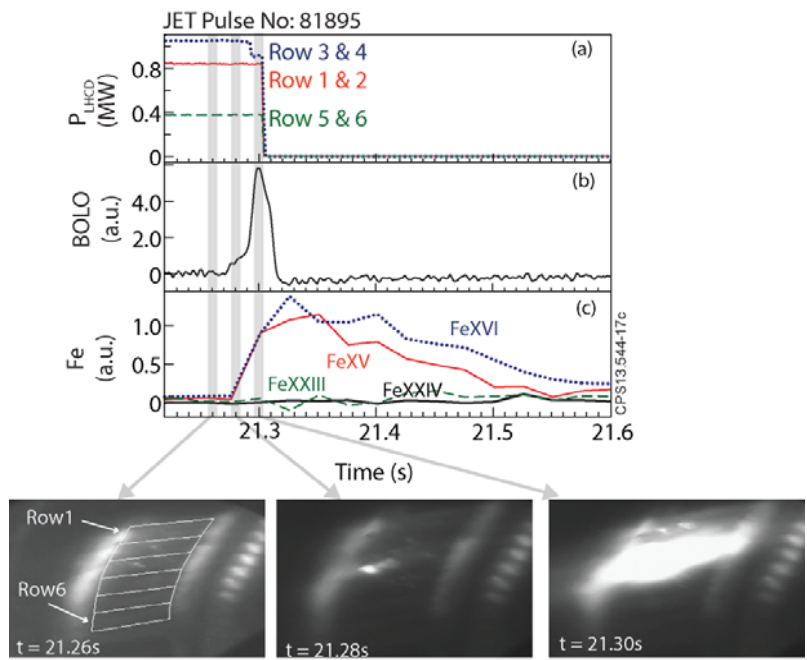

**Figure 17** Time evolution of (a) LH power per row, (b) bolometric signal for protection against arcs, (c) Fe line intensity (from mid-plane VUV spectroscopy) during an arc propagating on the LH structure.

## 4. Summary and Outlook

Overall the use of NBI, ICRF and LH systems in the Be/W environment was very successful with efficient heating and no damage linked to the use of the heating power. This was the consequence of a cautious approach which has provided, in parallel with the development of the ITER-like wall protection systems, confidence in the safe application of the power. So far there are no obstacles preventing further increase of the power for the high performance scenario development programme [51].

The NBI system that was upgraded in parallel with the change of wall reached a record power of 25.7 MW and a pulse length of 15 s (not at the same time) without any issue thanks to upgraded protection systems and to the new actively cooled duct. The use of W-coated PFC instead of Be in the main chamber shine-through areas combined with the newly implemented JET-ILW viewing systems monitoring areas at risk, allowed by the end of the campaign operation at same densities value than with the JET-C. It is foreseen that in the coming JET campaign the full NBI power resulting from the system upgrade will be reached and exploited.

Although all the private limiters of the ICRF antennas were replaced, no operational problem was observed. The heat-loads related of RF sheaths were found not to be threatening in the present condition but the high values observed in worst conditions (4.5 MW/m$^2$) justifies further monitoring by the PIW viewing system. Although leading to higher bulk radiation than with the JET-C, similar heating efficiency and plasma energy increase was found during ICRF with the JET-ILW. This higher radiation could be attributed mainly to W but also to Ni. The additional source of W during the ICRF is still unclear. Although interaction between high voltages regions in the RF sheaths and magnetically connected divertor outer tiles is likely, it was not experimentally measured. In parallel, W and Ni sources from main chamber recessed areas were evidenced. The W profiles in L-mode were found slightly hollow during the ICRF and peaked with the NBI. The presence of a higher W level towards the edge (compared to NBI) is likely to be mix of different effects: enhanced W source, strong sawtooth activity and transport effect due to high $T_e$ gradient although so far no firm conclusions are possible.

For the LHCD, the protective frame was changed to Be and a dedicated viewing system was installed. Because of the fear of arcing that could lead to disruption or hot-spot that could lead to Be melting, the power was kept to a relatively low level until the full protection system is develop. This should take place in the coming JET campaign. So far the power was increased were smoothly up to 2.5 MW (15 MW/m$^2$) without any specific impurity increase.

Note that many of the issues discussed here have already been addressed for ITER. For example, for the ITER 1MeV negative ion beams, the modeling shows that shine-through will not be an issue because of the long path length of the beam crossing the plasma [37]. For the ICRF, although it is critical to further understand any mechanisms involved in the process observed at JET, the first obvious step was taken by using antennas modeling tools (not available for the A2s) to minimize any residual parallel electric field [52].

**Acknowledgments**

This work, part-funded by the European Communities under the contract of Association between EURATOM/CCFE, was carried out within the framework of the European Fusion Development Agreement. For further information on the contents of this paper please contact publications-officer@jet.efda.org. The views and opinions expressed herein do not necessarily reflect those of the European Commission. This work was also part-funded by the RCUK Energy Programme under grant EP/I501045 .To obtain further information on the data and models underlying this paper please contact PublicationsManager@ccfe.ac.uk